\documentclass[aps,onecolumn,12pt,superscriptaddress,showpacs]{revtex4-1}

\usepackage{graphicx,color}
\usepackage{amsmath}
\usepackage{amssymb}
\usepackage{amsfonts}

\usepackage{bm}

\newcommand{\Name}[1]{{#1},}
\newcommand{\Book}[1]{\textit{#1}}
\newcommand{\Publ}[1]{({#1})}
\newcommand{\Year}[1]{{#1}}
\newcommand{\REVIEW}[4]{{\textit{#1},} {\textbf{#2}} {(#3)} {#4}}

\begin{document}

\title{Anisotropic Ginzburg-Landau and Lawrence-Doniach Models for Layered 
Ultracold Fermi Gases}

\author{Mauro Iazzi}
\affiliation{SISSA, Via Bonomea 265, I-34136 Trieste, Italy}

\author{Stefano Fantoni}
\affiliation{ANVUR, P.za Kennedy 20, I-00144 Roma, Italy}

\author{Andrea Trombettoni}
\affiliation{CNR-IOM DEMOCRITOS Simulation Center, Via Bonomea 265, I-34136
Trieste, Italy}
\affiliation{SISSA, Via Bonomea 265, I-34136 Trieste, Italy}


\begin{abstract}
We derive and study the anisotropic Ginzburg-Landau and 
Lawrence-Doniach models describing a layered superfluid ultracold Fermi gas in 
optical lattices. We compute from the microscopic model the Josephson couplings 
entering the Lawrence-Doniach model across the crossover BCS-BEC passing from 
the 3D isotropic case to the quasi-2D one, showing that a model with only 
nearest-neghbour Josephson coupling is not adequate at the unitary limit (since 
the pairs have a diameter larger than the interlayer distance). We also show 
that the effective anisotropy of the system is strongly reduced ad the unitary 
limit. Finally we obtain a relation between the interlayer Josephson couplings 
and the Ginzburg-Landau masses: we find that using only couplings between 
adjacent planes is correct in the BEC side, while at the unitary limit one has 
to use also next-nearest neighboring couplings.
\end{abstract}

\maketitle

\section{Introduction}
The presence and nature of the superfluid and superconducting transition in 
systems of different dimensionality is a widely studied and interesting topic. 
It is well known that the transition in two and three dimensions is radically 
different, as the first is governed by a Kosterlitz-Thouless mechanism of 
topological origin (the unbinding of vortices) and the second is the 
paradigmatic example of the symmetry breaking well described, at least 
qualitatively, by mean-field theories. This is one of the main reasons of the 
wide interest in the study of materials with properties in-between the two, 
i.e. layered superfluids and superconductors. Their phenomenological properties 
are usually well described by anisotropic effective models: a successful 
example is given by the Lawrence-Doniach (LD) model \cite{Lawrence1971} (in 
which the discrete layers are weakly coupled through Josephson terms) 
describing the electrodynamics of layered superconductors and copper oxides 
\cite{Tinkham1996}. Another related way to introduce the effect of the layering 
is to consider a Ginzburg-Landau (GL) theory having anisotropic masses 
(eventually phenomenologically determined) \cite{Tinkham1996}. A well known 
feature of these phenomenological models is the fact that they work well even 
if the underlying microscopic mechanisms are not known (as it occurs for 
high-$T_c$ cuprates). 

In this paper we study and compare the anisotropic GL and LD models describing 
a layered superfluid ultracold Fermi gas in optical lattices: the coefficients 
of these phenomenological models are derived from the microscopic underlying 
Hamiltonian, the parameters of which, e.g. the strength of the optical 
lattices, can be tuned with high accuracy in realistic experimental setups. We 
determine the mass tensor in the GL theory as a function of lattice anisotropy, 
filling and interaction, showing that near the unitary limit the effective 
anisotropy of the masses is significantly reduced. By tuning the strength of 
the optical lattices, the dimensionless anisotropy parameter $\gamma$ in 
realistic setups can vary in a wide range of values: $\gamma \sim 1-200$ (for 
comparison with layered superconductors, we remind that $\gamma \sim 7$ for 
YBCO and $\gamma \sim 150$ for BSCCO \cite{Tinkham1996}). We also derive the LD 
model, discussing a relation between the interlayer Josephson energy couplings 
in the LD model and the masses in the GL model.
shown to be equivalent to the anisotropic GL theory in the BEC limit, while in 
the BCS side 
next-nearest-neighbor couplings have to be taken into account, and a 

\section{Layered superconductors and superfluids} 
The first layered superconductor was realized alternating layers of graphite 
and alkali metals \cite{Hannay1965}, with very low critical temperatures 
($\approx0.1$K). It was followed by the discovery of naturally occurring 
compounds of transition-metal dichalcogenides layers (such as TaS$_2$) 
intercalated with organic, insulating molecules \cite{Gamble1970} and the 
creation of artificial samples with alternating layers of different metals, 
having a different transition temperature \cite{Ruggiero1980}, both systems 
with a higher critical temperature and thus more practical to study. The nature 
of the transition was found to be three dimensional \cite{Klemm1975}, and the 
systems were shown to be very well described by an anisotropic GL free energy 
at and above the transition, where the coherence length diverges. At lower 
temperatures where the length $\xi$ becomes comparable with the plane 
separation, the presence of a layer structure becomes important. To describe 
this Lawrence and Doniach proposed a tight-binding approximation 
\cite{Lawrence1971} able to describe the crossover between the 2D behaviors of 
essentially decoupled planes to the 3D. 
being the interlayer distance and $\xi_\perp$ the coherence length 
perpendicular to the planes). On this line the critical magnetic field parallel 
to the planes diverges in the LD model. To remove this divergence one has to 
include the effects of Pauli paramagnetism and spin-orbit coupling 
\cite{Klemm1975}.

Another very important class of superconductors having a layered structure is 
the one of cuprates, discovered in 1986 by Bednorz and M\"uller 
\cite{Bednorz1986}. 
copper oxide planes alternated to an insulating lattice with a perovskite-like 
structure. 
reach temperatures as high as $130 K$. 
Their layered structure makes the behavior of cuprates strongly anisotropic and 
the dimensional crossover temperature is in general very near the critical 
temperature for the resistive transition \cite{Tinkham1996}: despite the lack 
of a definitive consensus on the mechanism responsible for the 
superconductivity in the cuprates, it is by now established that the LD model 
successfully describe the transition and the electromagnetic response of 
cuprates \cite{Tinkham1996,Leggett2006}.

The study of the dimensional crossover in cuprates is not in general 
an easy task: on one hand, the smallest barrier attainable is that of a single 
plane of an insulating material, on the other hand inserting several planes is 
a delicate process as the separation can become nonuniform 
\cite{Leggett2006}. From this point of view, ultracold atoms (and in particular 
ultracold fermions) in optical lattices provide a favorable setting: the 
tunneling rate $t$ can be controlled by adjusting the optical lattice 
parameters \cite{Morsch2006,Bloch2008} and the on-site energy coefficient $U$ 
can be controlled through Feshbach resonances 
\cite{Bloch2008}. The ratio $U/t$ can be then controlled with high precision, 
as well as the geometry and the anisotropy of the system. 
respect to all other energy scales), the gas in the center is well described by 
an essentially homogeneous model.
For ultracold bosons, optical lattices have been used to induce a 
Mott-insulator/superfluid quantum phase transition \cite{Greiner2002} and to 
study Josephson dynamics in many- and two- wells settings 
\cite{Cataliotti2001,Albiez2005}. 
The study of ultracold fermions in optical lattices also stimulated a huge 
amount of experimental work 
\cite{Modugno2003,Chin2006,Stoferle2006,Schneider2008,Jordens2008,Sommer2011}. 
The low energy properties of ultracold fermions in optical lattices are 
described by Hubbard-like models \cite{Hofstetter2002,Titvinidze2010}:
in presence of attractive interaction, with deep optical lattices 
one then can have   
a physical realization of the attractive Hubbard model 
\cite{Singer1996,Sewer2002,Toschi2005,Burovski2006,Chien2008}. 

Superimposing a 1D optical lattice to an ultracold Fermi gas gives raise to a 
stack of two-dimensional layers. Several equilibrium and dynamical properties 
of a Fermi superfluid in presence of a superimposed 
optical lattice have been theoretically studied 
\cite{Orso2005,Iskin2009,Fialko2010,Watanabe2011}. 
Very recently, the evolution of fermion pairing from 3D to 2D was 
experimentally studied, showing the opening of a gap in the radiofrequency 
spectrum when the dimensionality is reduced \cite{Sommer2011}. The confinement 
of Fermi mixtures in 2D was also experimentally investigated 
\cite{Gunter2005,Du2009,Martiyanov2010,Frohlich2011,Dyke2011,Feld2011}, 
showing properties different from the 3D case \cite{Giorgini2008}. 
The observation of a pairing pseudogap in a 2D Fermi gas, using 
momentum-resolved photoemission spectroscopy, 
has been reported \cite{Feld2011}. 

An important advantage of the use of optical lattices is that 
when it is applied (say along the $z$ direction) 
then the interlayer coupling $t_\parallel$ between 2D pancakes along $z$ can be 
easily modified by varying the strength of the optical lattice. Furthermore, if 
two additional optical lattices are added in the $xy$ plane, then the tunneling 
$t_\perp$ can be tuned separately from $t_\parallel$ and the layering can be 
tuned by varying the ratio $t_\parallel/t_\perp$. 

\section{Effective models} 
Near the critical temperature, the superfluid wavefunction $\Psi$ is small, so 
that the free energy difference $\delta F$ between the normal and superfluid 
state can be expanded in powers of $|\Psi|^2$ \cite{Ginzburg1950}: for a 
layered superconductor/superfluid $\delta F$ reads then
$$
\delta F[\Psi] = \int d\vec{r} 
\Big\{\frac{\hbar^2}{2m_\parallel}(|\partial_x\Psi|^2+|\partial_y\Psi|^2) +
$$
\begin{equation} \label{eq:GL}
+ \frac{\hbar^2}{2m_\perp}|\partial_z\Psi|^2 + \alpha(T) 
|\Psi|^2 + \beta |\Psi|^4\Big\}.
\end{equation}
The phenomenological coefficients $\alpha$ and $\beta$ appearing 
in~(\ref{eq:GL}) can be derived in general from experiments or from the 
microscopic Hamiltonian; the masses $m_{\parallel,\perp}$ are those of the 
Cooper pairs. 
minimizing~(\ref{eq:GL}) with respect to $\Psi$. 
state, 
hence 
In the LD tight-binding approach one has a 2D superfluid wavefunction 
$\Psi_n(x, y)$ in the $n$th plane along $z$ and the term with transverse 
derivative becomes 
$-J_1\sum_n|\Psi_n(r)-\Psi_{n+1}(r)|^2$, with $J_1$ being the nearest-neighbor 
tunneling coefficient \cite{Tinkham1996}.
spacing of the periodic potential along $z$).
model from the 
lattice.

The effective Hamiltonian describing a two-component Fermi 
gas in a layered optical lattice reads 
$\hat{H}-\mu \hat{N}$, with the Hamiltonian $\hat{H}$ given by
\begin{equation} \label{eq:Hubbard}
\hat{H} = -\sum_{\langle ij \rangle\sigma}
(t_{ij} \hat{\phi}^{\dagger}_{i\sigma}\hat{\phi}^{\vphantom{\dagger}}_{j\sigma} 
+ h.c. ) - 
U\sum_{i}
\hat{\phi}^{\dagger}_{i{\uparrow}}
\hat{\phi}^{\dagger}_{i{\downarrow}}
\hat{\phi}^{\vphantom{\dagger}}_{i{\downarrow}}
\hat{\phi}^{\vphantom{\dagger}}_{i{\uparrow}}, 
\end{equation}
where $\hat{N}=\sum_{i\sigma}\hat{\phi}^{\dagger}_{i\sigma}
\hat{\phi}^{\vphantom{\dagger}}_{i\sigma}$ is the number operator, $\mu$ is the 
chemical potential, $i,j$ are site in dices, $\sigma={\uparrow},{\downarrow}$ 
is the spin index and $\hat{\phi}^{\dagger}_{i\sigma}$ is the fermionic 
operator (the sum is on distinct pairs of nearest-neighbors). Furthermore 
$t_{ij}$ is the hopping rate between the sites $i$ and $j$, taken to be 
equal to $t_\parallel$ in the $xy$ planes and $t_\perp$ along the $z$ 
direction. 
The isotropic 3D case corresponds to $t_\parallel=t_\perp$, while the isotropic 
2D case to $t_\perp=0$ (isolated pancakes). 
The single-particle energies of Hamiltonian~(\ref{eq:Hubbard}) for $U=0$ are 
$\epsilon_{\vec{k}}^{(0)}= -2t_\parallel(\cos{k_xd}+\cos{k_yd})-2t_\perp 
\cos{k_zd}$ with total bandwidth 
\begin{equation}\label{total_D}
D = 8t_\parallel+4 t_\perp.
\end{equation} 
The interaction is assumed attractive, $U>0$, and the total number of fermions 
per site is denoted by $n$. 
Estimates of the value of $U$ for which the BCS-BEC crossover occurs gives - 
(in the isotropic case $t_\parallel=t_\perp$) 
$U \approx 0.7D$ in 3D and $U ~ \approx 0.2 D$ in 2D 
\cite{Singer1996,Sewer2002,Toschi2005,Burovski2006,Chien2008}. 

The physical system modeled by~(\ref{eq:Hubbard}) has 
two optical lattices in the $x,y$ directions 
(determining the tunneling rate $t_\parallel$  between the sites in the $xy$ 
planes) and a different lattice in the $z$ direction, with interlayer coupling 
$t_\perp$ 
between neighbouring planes. 
In our subsequent determination of the coefficients of the anisotropic GL and 
LD models the crucial ingredient is to have the optical lattice along $z$ and 
the mechanism of layering in the form of tunneling between adjacent layers: if 
the optical lattices in the $x,y$ directions are absent (or small), then the 
anisotropic 
GL~(\ref{eq:GL}) would be again retrieved (with $m_\parallel=2m$, where $m$ is 
the mass of the fermionic atoms). 
Similarly, in writing~(\ref{eq:Hubbard}) we assumed in each minimum of the 
lattice there is a total number of fermions $n \le 2$ 
(i.e., only the lowest band of the periodic potential is occupied): otherwise, 
we would have a multi-band Hubbard model, 
but again a free energy of the form~(\ref{eq:GL}) would be retrieved. 

\section{Estimates of parameters in realistic potentials} 

\begin{figure}[t]
\begin{center}
\begin{picture}(200,130)
{\includegraphics[scale=0.25]{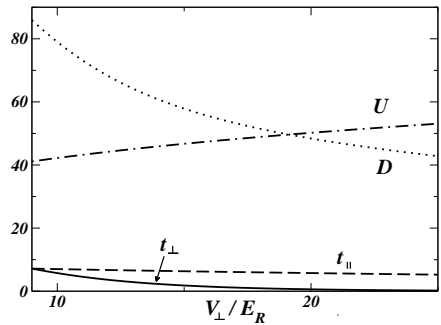}}
\end{picture} 
\caption{Values in nanokelvins 
of the parameters $t_\perp$ (solid line), $t_\parallel$ (dashed) and 
$U$ (dot-dashed) as a function of $V_\perp/E_R$; 
the total bandwidth $D=8 t_\perp+4t_\parallel$ (dotted) is 
reported as well. Estimates are for $^6$ Li atoms (the parameter $U$ is 
computed for $| a | =300 a_0$). The parameter $V_\parallel$ is fixed to be 
$V_\parallel=9E_R$.}
\label{exp_par}
\end{center}
\end{figure}

The hopping rates $t_\perp$, $t_\parallel$ and the interaction 
coefficient $U$ are as usual expressed as integrals of Wannier functions 
\cite{Morsch2006}. We present here estimates for a 
realistic setup considering  
an optical lattice potential 
$V_{opt} \left( \vec{r} \right)=V_\perp \left[ \cos^2{(k x)} + \cos^2{(k y)} 
\right] + V_\parallel \cos^2{(k z)}$: the parameters $V_\perp$, $V_\parallel$ 
can be controlled with the power of the 
laser beams in the $\perp=x,y$ and $\parallel=z$ directions: the minima 
of different wells are separated by an energy barrier $V_\perp$ 
($V_\parallel$) along the $x,y$ ($z$) directions. The spacing 
of the lattice, supposed for simplicity equal in the three directions, is 
$d=\lambda/2$ where $\lambda=2 \pi/k$; energies 
are usually defined in units of the recoil energy $E_R=\hbar^2k^2/2m$ 
\cite{Morsch2006}.
Deep lattices are characterized by having 
the chemical potential (much) smaller  
than $V_\perp$, $V_\parallel$. Typical values for the parameters are 
$\lambda \sim 1 \mu m$ and $V_\perp, V_\parallel \sim 1 - 10kHz$; 
for larger 
values of $\lambda$ (say $\lambda \sim 10 \mu m$ as in \cite{Albiez2005}) 
values larger than $ 500 Hz$ are enough to have tunneling dynamics within 
the first band. 
For $^6$ Li atoms 
(using a scattering length $| a | =300 a_0$), 
for $V_\parallel=5E_R$ and $V_\perp=3.6 V_\parallel$ one gets $U/D \approx 
0.6$, 
$t_\perp / t_\parallel \approx 0.15$ and $D/k_B \approx 60nK$. For $V_\parallel=
9E_R$ and $V_\perp=1.25 V_\parallel$ one gets $U/D \approx 0.6$, 
$t_\perp/t_\parallel \approx 0.6$, $D/k_B \approx 60nK$. 
In Fig.\ref{exp_par} we plot 
the coefficients $t_\perp$, $t_\parallel$, $U$ vs. $V_\perp$ 
for a set of realistic values of the optical lattice parameters.

\begin{figure}[t]
\begin{picture}(200,130)
\put(0,-10){\includegraphics[scale=0.6]{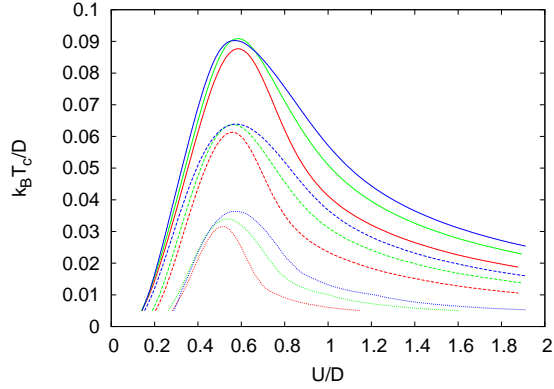}}
\end{picture} 
\caption{Critical temperature $k_B T_c$ vs. $U$ 
(both scaled in units of 
$D=8t_\parallel+4t_\perp$) at various particle densities - 
$n=0.9$ (solid lines), $n=0.5$ (dashed lines), $n=0.1$ (dotted lines) - 
and anisotropies $t_\perp/t_\parallel= 1,\ 0.3,\ 0.1$ (in each set from top to 
bottom).}
\label{crit_temp}
\end{figure}

\section{Critical temperature and chemical potential}
Since the coefficients of~(\ref{eq:GL}) depend on the critical temperature 
$T_c$ and the chemical potential $\mu$ - which in turn depend on the 
parameter of the microscopic Hamiltonian~(\ref{eq:Hubbard}) - we preliminary 
study $T_c$ and $\mu$ as a function of the interaction $U$, the layering 
$t_\perp/t_\parallel$ and the density $n$.

To derive the coefficients of the anisotropic GL we proceed in a 
path-integral setting \cite{SaDeMelo1993,Randeria1995,Zwerger2011}: 
the fermionic atoms are described by Grassmann fields. 
First, the Hubbard-Stratonovich transformation is performed;
the fermionic degrees of freedom are then integrated out obtaining the bosonic
effective action: performing the saddle approximation leads to the mean-field
equations. At finite temperature, far from the deep BCS regime 
(which is obtained for $U \ll D$), one has
to consider the effects of higher order fluctuations \cite{Randeria1995} and
solve two coupled equations one for the gap, the other enforcing 
the conservation of the number of particles.
The first equation is the saddle point
equation for the order parameter $\Delta_0$:
\begin{equation}\label{eq_1}
\frac{1}{U} = \sum_{\vec{k} \in BZ} \frac{\tanh(\beta
E_{\vec{k}}/2)}{2E_{\vec{k}}}, 
\end{equation}
where $E_{\vec{k}} = \sqrt{\epsilon_{\vec{k}}^2+\Delta_0^2}$ and 
$\epsilon_{\vec{k}}=\epsilon_{\vec{k}}^{(0)}-\mu$. The second equation is the
number equation  
\begin{equation}\label{eq_2}
n = -\frac{\partial}{\partial\mu}F
\end{equation}
where $F$ is the free energy, which at the 
critical temperature (defined by $\Delta_0 \equiv 0$) 
is given by $F/k_B T_c= -\sum_{\vec{k} \in BZ} 
\ln{\left(1+e^{-\beta\epsilon_{\vec{k}}} \right)} + \sum_{q} 
\ln{\left\{ \frac{1}{U}+\chi(q) \right\}}$
where $\chi(q)=k_B T_c \sum_{k} G^+_k G^-_{k+q}$ 
is the pair polarization function with $G_{k}^{\pm}=
\left(i \omega \pm \epsilon_{\vec{k}} \right)/(\omega^2+
\epsilon_{\vec{k}}^2)$ (we used the notation
$k=(i\omega;\vec{k})$ where the wavevectors $\vec{k}$ belong to the Brillouin
zone (BZ) and the $\omega_\ell$ are the Matsubara
frequencies).

To determine the critical temperature one has to solve 
Eqs.~(\ref{eq_1})-(\ref{eq_2}) with $\Delta_0=0$: 
these equations 
involves integrals in $6$ variables, evaluated first 
finding the poles of the integrand and analytically integrating around them 
and then integrating numerically \cite{Iazzi2012}.
In this way one obtains $T_c$ and
$\mu$ as a function of the tunneling rates $t_\parallel$ and $t_\perp$, the
particle density $n$ and the interaction $U$. As an example, 
for $^6$Li atoms and for $V_\parallel=9E_R$ and $V_\perp=1.25 V_\parallel$ 
one has $T_c \approx 5nK$ 
($T_c / T_F \approx 0.3$).  
The results for the critical temperature $T_c$ for various values 
of filling $n$ and of the ratio $t_\perp/t_\parallel$ 
are plotted in Fig.\ref{crit_temp}: we plot both $T_c$ and $U$ in units of 
$D$, showing that in these units the effect of the anisotropy 
is relatively small (the critical temperature in the interval 
$0.5  < t_\perp/t_\parallel < 1$ is very similar 
to the $T_c$ of the isotropic case). 

\section{Coefficients of anisotropic GL and LD models}
The coefficients of the GL free energy~(\ref{eq:GL}) are found to be 
$$
\frac{\hbar^2}{2M_a} = 
\Bigg[\sum_{\vec{k}}\Big\{\beta_c^2 XY+\frac{\beta_c Y}{\epsilon_{\vec{k}}}-
\frac{2X}
{\epsilon_{\vec{k}}^2}\Big\}\frac{(\partial_{q_a}\epsilon_{\vec{k}+\vec{q}})^2}{
8\epsilon_{\vec{k}}} +
$$
\begin{equation}\label{eq_m}
+\sum_{\vec{k}} \left\{\frac{X}{4\epsilon_{\vec{k}}^2} - \frac{\beta_c Y}
{8\epsilon_{\vec{k}}}\right\}\partial_{q_a}^2
\epsilon_{\vec{k}+\vec{q}}\Bigg]_{\vec{q}=0},
\end{equation}
\begin{equation}
\alpha = \frac{1}{U}-\sum_{\vec{k}}\frac{X}{2\epsilon_{\vec{k}}}, \,\,\,
\beta = \sum_{\vec{k}} \left\{\frac{X}{4\epsilon_{\vec{k}}^3} - 
\frac{\beta_c Y}{8\epsilon_{\vec{k}}^2}\right\}:
\label{alpha_beta}
\end{equation}
in~(\ref{eq_m}) $a=x,y,z$ denotes the direction and the mass parameters $M_a$  
are related to the masses in the GL 
energy~(\ref{eq:GL}) by $M_a=m_a U^2$ 
(with $m_x=m_y\equiv m_\parallel$, $m_z \equiv m_\perp$). 
Furthermore it is 
$X=\tanh(\beta_c \epsilon_{\vec{k}}/2)$ and 
$Y=\cosh^{-2}(\beta_c \epsilon_{\vec{k}}/2)$ with $\beta_c=1/k_BT_c$. 
Eqs.~(\ref{eq_m})-(\ref{alpha_beta}) are 
obtained expanding the bosonic 
action up to the fourth order and evaluating it at the minimum.
 
The coefficients of~(\ref{eq:GL}) depend on the chemical potential 
$\mu$ and the critical temperature $T_c$, which are in turn derived 
minimizing the free energy evaluated by integrating the gaussian 
(i.e., at one-loop level)  fluctuations 
above the saddle point of the path integral expression for the 
partition function. 
Once that $\mu$ and $T_c$ are determined, they are inserted in the 
coefficients of the GL~(\ref{eq:GL}). To account 
for thermal fluctuations one should keep the GL form of the effective action 
and compute observables using the path-integral instead of minimizing  
the action \cite{Larkin2005}. Thermal fluctuations play 
an important role 
around the unitary limit and in the strongly anisotropic limit, where 
the behavior of the system is essentially governed by 2D fluctuations. 
In particular we expect that our computation is not correct 
in the pure 2D case, $t_\parallel=0$, and that the range of temperature 
close to $T_c$ at which Eqs.~(\ref{eq_m})-(\ref{alpha_beta}) are valid 
becomes smaller for $t_\parallel \to 0$.
The validity (and methods of solutions) of the GL equations for 
2D and layered superconductors, also in presence of magnetic fields, 
has been discussed \cite{Ullah1991,Tesanovic1994,Benfatto2000,Murray2010}: 
for ultracold fermions, the comparison with 
experiments in quasi-2D setups 
will possibly 
shed light on the quantitative validity 
of the GL~(\ref{eq:GL}). 

\begin{figure}[t]
\begin{picture}(200,130)
\put(-10,0){\includegraphics[scale=0.6]{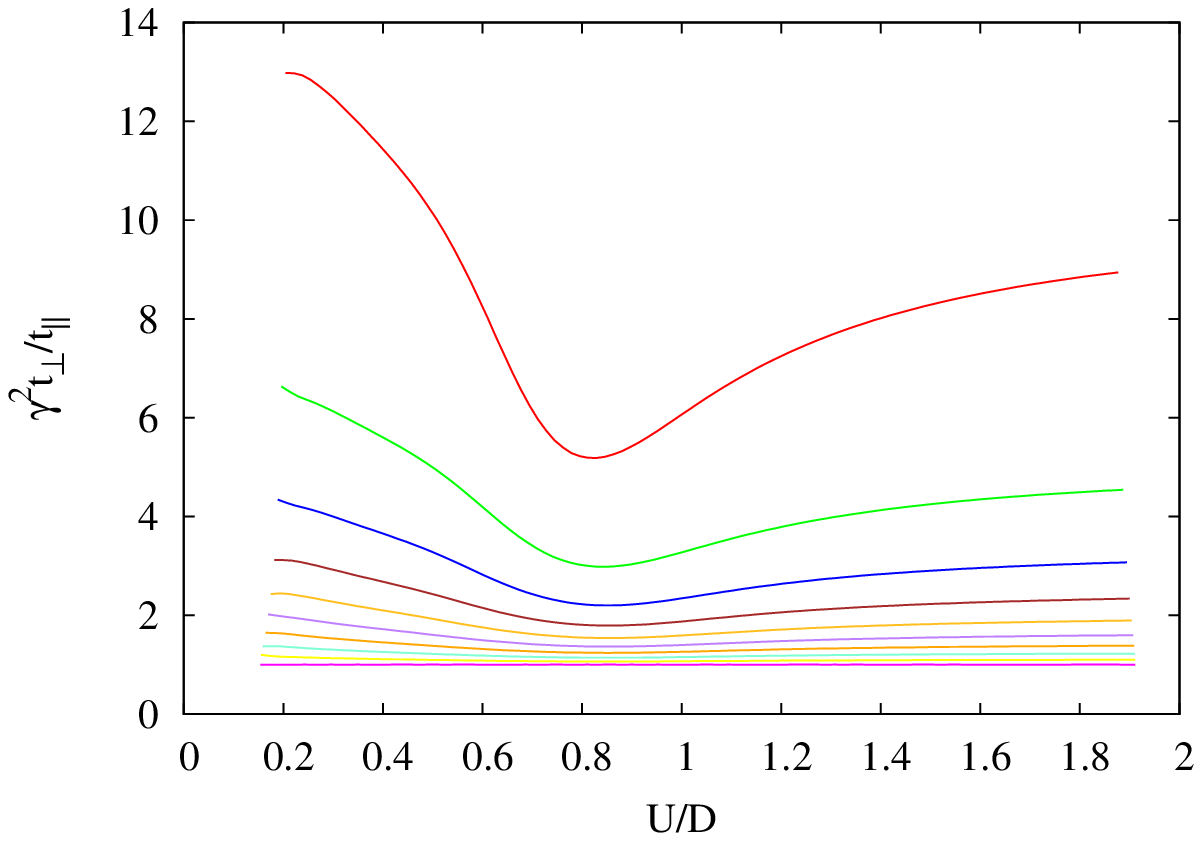}}
\put(105,83){\colorbox{white}{\includegraphics[scale=0.25,clip=true,trim=25 20 
10 10]{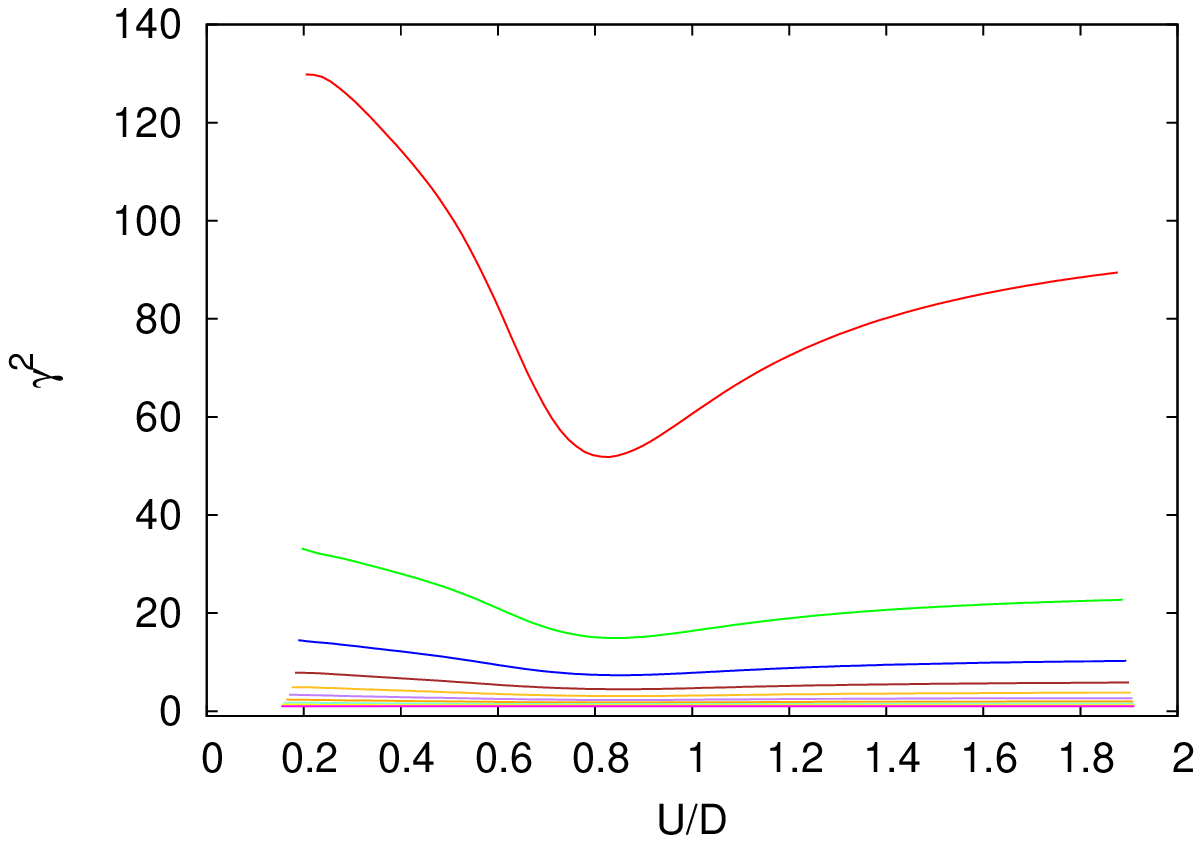}}}
\end{picture} 
\caption{Effective anisotropy $\gamma_{eff}^2=m_\perp t_\perp / 
m_\parallel t_\parallel$ as a function of $U/D$
for different anisotropies (at $n=0.5$): from top to bottom 
$t_\perp/t_\parallel=0.1$, $0.2$, $0.3$, $0.4$, $0.5$, 
$0.6$, $0.7$, $0.8$, $0.9$, $1$. Inset: Anisotropy parameter 
$\gamma^2=m_\perp / m_\parallel$ vs. $U/D$ for the same values 
of anisotropies $t_\perp/t_\parallel$ and $n$.}
\label{anisotropy}
\end{figure}

\begin{figure}[t]
\begin{picture}(200,130)
\put(0,-10){\includegraphics[scale=0.6]{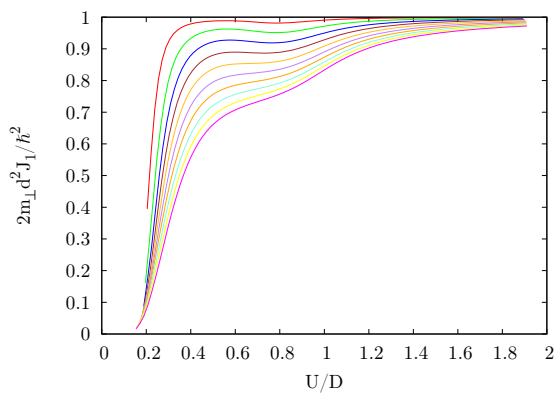}}
\end{picture} 
\caption{Plot of the quantity $2m_\perp d^2 J_1 / \hbar^2$ vs. $U/D$ 
at $n=0.5$ (with $d$ lattice spacing) 
for the same values of  $t_\perp/t_\parallel$ of Fig.\ref{anisotropy}.} 
\label{comp_LD_GL}
\end{figure}


It is also possible to derive the LD tunneling coefficients: in the LD model, 
the $xy$ layers are weakly coupled along the $z$ direction through 
Josephson terms. Usually, only Josephson couplings between adjacent planes are 
considered. Here we consider the more general possibility of longer range 
tunneling: 
this amounts to consider a Josephson contribution $-\sum_{n,m} J_{n,m} \mid 
\psi_n-\psi_m \mid^2$ to the energy, where $n,m$ are denoting the different 
planes.
We find
\begin{equation}
J_{n,m}=-\frac{U^2 d}{2 \pi} \, \int dq_z e^{-i(n-m)q_zd} \chi(i\omega=0;0, 0, 
q_z)
\label{Jnm}
\end{equation}
where $\chi(q)=\chi(i\omega;\vec{q})$ is the pair polarization function 
evaluated at the critical point and reported after 
Eq.~(\ref{eq_2}). 
In presence of lattices in the $x,y$ directions, 
one can similarly define Josephson couplings $J_{n,m}^{\parallel}$. 
It is possible to determine a useful relation between the coefficients 
$J_{n,m}$ and the mass $m_\perp$ of the anisotropic GL model: one finds 
\begin{equation}
\frac{\hbar^2}{2m_\perp d^2} = \sum_{p=1}^{\infty} p^2 J_p,
\label{Jnm_id}
\end{equation} where we used the notation $J_p \equiv J_{n,n+p}$ 
(we are assuming translational invariance along $z$): then $J_1 = J_{n,n+1}$.

\section{Results} 
Using the solution for $T_c$ and $\mu$ of Eqs.~(\ref{eq_1})-(\ref{eq_2}) 
we can use Eq.~(\ref{eq_m}) to evaluate the masses 
$m_\parallel$, $m_\perp$ and the mass 
anisotropy parameter $\gamma$ usually defined \cite{Tinkham1996} as   
\begin{equation}
\gamma^2 = \frac{m_\perp}{m_\parallel}.
\label{gamma}
\end{equation}

The results for the anisotropy parameter $\gamma$ of a 
layered ultracold Fermi gas 
are shown in the inset of Fig.\ref{anisotropy}: 
$\gamma$ significantly decreases close to the unitary limit due to the effect 
of the increase of $T_c$ in the crossover region. 
This result is consistent with the qualitative expectation that the effect 
of external anisotropies is softened in the unitary limit. We investigated 
the behaviour of $\gamma$ vs. filling, finding that the filling modifies 
$\gamma$ on the BCS side, while $\gamma$ 
on the BEC side remains virtually identical. To understand if and how much  
the collective behavior enhances or not the (external) anisotropy 
artificially tuned by imposing a layering through a value of 
$t_\parallel/t_\perp$ smaller than one, we defined an {\em effective} 
anisotropy defined as 
\begin{equation}
\gamma_{eff}^2=\gamma^2 \,
\frac{t_\perp}{t_\parallel}=\frac{m_\perp}{m_\parallel} \, 
\frac{t_\perp}{t_\parallel}.
\label{gamma_eff}
\end{equation}
The reason behind the definition~(\ref{gamma_eff}) is that one 
expects that the effective mass along the direction $a$ is $\propto 1/t_a$. 
Then a value of $\gamma_{eff}$ larger than one signals an effective
enhancement of the anisotropy (with respect to the ``external'' anisotropy
$t_\parallel/t_\perp$). The results are drawn in
Fig.\ref{anisotropy}: 
it is seen that in the BCS side the enhancement 
is more pronounced with respect to the BEC side. Again, the enhancement is 
reduced in the unitary limit. 
We also checked that in the BEC limit 
the quantity $\gamma^2 \cdot (t_\perp/t_\parallel)^2$ approaches $1$ for 
different fillings: a detailed study of $\gamma$ and $\gamma_{eff}$, as well 
as of the coefficients $\alpha$ and $\beta$, as a function 
of the layering across the crossover will be presented elsewhere. 

Finally, we numerically evaluated the LD coefficients~(\ref{Jnm}) 
and compared them to the GL masses in Eq.~(\ref{eq_m}). 
We found that in the BEC side the nearest-neighbor Josephson coupling 
$J_1$ is enough to give the correct value $m_\perp$: 
the numerical estimate of both $m_\perp$ and the $J_p$'s show that 
in the BEC side the identity~(\ref{Jnm_id}) is very well saturated 
only by $J_1$, i.e. only the $p=1$ term gives an 
important contribution: therefore in the BEC side 
one has $J_1 \approx \hbar^2/2m_\perp d^2$.

However, in the BCS side and at the unitary limit 
our results show that the higher order $J_p$'s (with $p \ge 2$) 
are not negligible, 
as shown in Fig.\ref{comp_LD_GL} where we plot the quantity 
$2m_\perp d^2 J_1 / \hbar^2$ vs. $U$: according~(\ref{Jnm_id}), when this 
quantity is close to $1$, using only $J_1$ is enough to adequately describe the 
superfluid. A careful numerical investigation 
of the interlayer couplings $J_p$ shows at the unitary limit $J_2$ is roughly 
enough to saturate identity~(\ref{Jnm_id}), also in the isotropic limit. 
As example, for $U/D=0.7$, i.e. approximately where there is the unitary limit, 
one has for $n=0.5$ 
and $t_\perp / t_\parallel=0.6$ 
that $J_2/J_1 \approx 0.05$; in this case, 
the identity~(\ref{Jnm_id}) is satisfied 
up to $70\%$ with only $J_1$ and up to $95\%$ with $J_1$ and $J_2$. 
This not the case in the BCS side, where other $J_p$'s are needed: 
however, the deep BCS regime 
is currently out of reach of experiments since the corresponding critical 
temperature $T_c$ is very low.

The reason for having non-negligible $J_2$ 
(and eventually higher order $J_p$) is due to the fact 
that only in the BEC side the size of the pairs is not larger than 
than the lattice spacing: far form the BEC side the 
LD model with only nearest-neighbor Josephson 
couplings is not adequate to describe the superfluid 
(even though one could think to determine  
an optimized $J_1$ also far from the BEC side and have an effective LD model 
with only nearest-neighbour tunneling). 
Fig.\ref{comp_LD_GL} also shows 
a non-monotonous behavior 
around the unitary limit and 
that the validity region of the LD model with only 
Josephson couplings $J_1$ along $z$ is seen to 
increase when the layering becomes stronger 
(i.e., $t_\perp/t_\parallel$ becomes smaller). 

p

\section{Anisotropy parameter and effective masses}
The anisotropy parameter $\gamma$ plays a crucial role in the electrodynamics 
of layered superconductors since a formula giving the value of a quantity for a 
general anisotropy parameter $\gamma$ given 
its value for $\gamma=1$ holds \cite{Blatter1992}: the 
case of the dependence of the upper critical magnetic field upon $\gamma$ 
has been discussed \cite{Mineev2001}. 
Values of $\gamma$ for several superconducting compounds 
have been determined from experimental data \cite{Tinkham1996}. 

In layered superconductors, the effects and the penetration of a magnetic field 
have been studied \cite{Tinkham1996,Rosenstein2010}. They can be incorporated 
via the minimal substitution 
$\partial_a\rightarrow\partial_a-i\frac{e}{c}A_a$. 
$F_H=(\vec{B}-\vec{H})^2/8\pi$. 
and the response of a layered superconductor to an electromagnetic field have 
been an established field of research for decades 
\cite{Tinkham1996,Blatter1992,Rosenstein2010}. 
For 
ultracold atoms one can simulate a fictitious magnetic field realized by 
by rotating the trapping potential \cite{Fetter2009} together 
with the lattice \cite{Tung2006,Williams2010} or by 
using spatially dependent optical couplings between internal 
states of the atoms giving raise to a synthetic field \cite{Lin2009}. By using 
a Fermi mixture in optical lattices subjected to a synthetic magnetic field 
it is possible then to reproduce the phenomenology of the electrodynamics of 
layered superconductors in the context of ultracold atoms.

In layered Fermi superfluids, the value of $\gamma$ affects the critical value 
of a  synthetic magnetic field (the atomic counterpart of a magnetic field on a 
layered superconductor \cite{Tinkham1996}). 
Supposing to have an effective magnetic field $B$ around the axis $a$, 
at a certain critical value $B_{c2; a}$ the density profile 
of the cloud will show that the vortices have melt: this could 
be observed by the
disappearance of the vortex cores from the images of the cloud 
\cite{Fetter2009}. These critical rotations obey the relation 
$B_{c2 \parallel} / B_{c2\perp}=\gamma$. 

We finally observe that in experiments with ultracold atomic clouds 
the effective masses, and 
then the mass ratio~(\ref{gamma}), could be experimentally estimated 
from the measurement of the frequency of the collective excitations.
E.g., suddenly moving an external parabolic trap 
along the $a$ direction, dipole 
oscillations are induced: the normal 
(i.e., non-superfluid) part of the cloud would remain on a side of the 
harmonic potential and eventually roll down up to the center (this 
since there is an optical lattice in the $a$-direction). However, 
the superfluid part would oscillate around the center. From the 
oscillations one can measure the effective masses $m_a$: 
indeed one has that the frequency without 
the lattice is $\omega_a$ and with the lattice is $\omega_a \sqrt{m/m_a}$ 
(this experiment has been already realized 
with ultracold bosons at finite temperature: 
see e.g. Fig.2(b) of \cite{Cataliotti2001}).

\section{Conclusions} In this paper we studied effective models for 
equilibrium properties of the superfluid phase of an ultracold layered Fermi 
gas in the presence of an optical lattice near the critical temperature. This 
system is described by an anisotropic Ginzburg-Landau theory (that is also used 
for other systems such as layered superconductors). We derived 
and studied the coefficients of the Ginzburg-Landau equation: 
the anisotropy was shown to be enhanced by the superfluid, 
and the anisotropy parameter $\gamma$ decreases near the crossover region. 
The dependence on filling was found to be important on the BCS side of the 
crossover and negligible in the BEC side. Possible experimental measurements 
of the effective masses and the anisotropic parameter have been 
as well discussed.
We also derived the Lawrence-Doniach model for the 
layered ultracold Fermi gas and we discussed 
a relation between the interlayer Josephson couplings in the Lawrence-Doniach 
model 
and the masses in the Ginzburg-Landau energy. We found that using 
only couplings between adjacent planes is correct in the BEC side: 
at the unitary limit one can limit oneself to the use of only 
$J_1$ and $J_2$, while in the BCS side 
contributions from longer range interlayer couplings appear. 

The obtained results link the underlying microscopic description 
of a layered ultracold Fermi gas 
to the macroscopic properties of the superfluid and 
show that tuning the experimentally controllable parameters 
such as interaction strength, filling and lattice strength it is possible 
to emulate the phenomenological properties of layered superconductors.

\acknowledgments
Discussions with K. Schmidt and G. Roati are gratefully acknowledged.

\end{document}